\pdfoutput=1
\documentclass[letterpaper, 10 pt, conference]{ieeeconf} 
\IEEEoverridecommandlockouts                              

\usepackage{graphicx}
\usepackage{graphics} 
\usepackage{amsmath} 
\usepackage{amssymb}  
\usepackage{wrapfig} 
\usepackage{cleveref} 
\usepackage{cite}
\usepackage{graphicx} 
\usepackage{dblfloatfix} 
\usepackage{soul,color}

\usepackage[caption=false,font=footnotesize]{subfig}

\def\mf{\mathbf}

\def\mc{\mathcal}

\def\zd{\mf z_D}
\def\za{\mf z_A}
\def\zb{\mf z_B}

\def\xd{\mf x_D}
\def\xa{\mf x_A}

\def\taa{\tau_A(\za,\zb)}
\def\tdd{\tau_D(\zd,\zb)}

\def\bql{\begin{equation}}
\def\eql{\end{equation}}

\def\pp{p(\zd,\za,\zb)}

\Crefname{figure}{Fig.}{Figures}
\crefname{figure}{Fig.}{Figures}
\crefname{table}{Table}{Tables}
\crefformat{table}{Table~#1}
\crefformat{equation}{equation (#1)}
\Crefformat{Equation}{Equation (#1)}
\crefformat{algorithm}{Algorithm~#1}  

\title{\LARGE \bf
Learning Decentralized Strategies for a Perimeter Defense Game with Graph Neural Networks}

\author{Elijah S. Lee$^{1}$, Lifeng Zhou$^{2}$, Alejandro Ribeiro$^{1}$, and Vijay Kumar$^{1}$
\thanks{We gratefully acknowledge the support from 
ARL DCIST CRA under Grant W911NF-17-2-0181,
NSF under Grants CCR-2112665, CNS-1446592, and EEC-1941529,
ONR under Grants N00014-20-1-2822 and N00014-20-S-B001, 
Qualcomm Research, 
NVIDIA,
Lockheed Martin,
and C-BRIC, a Semiconductor Research Corporation Joint University Microelectronics Program program cosponsored by DARPA.
}
\thanks{$^{1}$The authors are with the GRASP Lab, University of Pennsylvania, Philadelphia, PA 19104, USA. 
{\tt\footnotesize \{elslee, aribeiro, kumar\}@seas.upenn.edu}
}%
\thanks{$^{2}$ The author is with the Department of Electrical and Computer Engineering, Drexel University, Philadelphia, PA 19104, USA. {\tt\footnotesize email: {lz457}@drexel.edu.}
}%
}

\begin{document}

\maketitle
\thispagestyle{empty}
\pagestyle{empty}

\begin{abstract}
We consider the problem of finding decentralized strategies for multi-agent perimeter defense games. In this work, we design a graph neural network-based learning framework to learn a mapping from defenders' local perceptions and the communication graph to defenders' actions such that the learned actions are close to that generated by a centralized expert algorithm. We demonstrate that our proposed networks stay closer to the expert policy and are superior to other baseline algorithms by capturing more intruders. Our GNN-based networks are trained at a small scale and can generalize to large scales. To validate our results, we run perimeter defense games in scenarios with different team sizes and initial configurations to evaluate the performance of the learned networks.
\end{abstract}

\section{Introduction}
\label{sec:intro}


The perimeter defense game, as a variant of the pursuit-evasion game~\cite{isaacs1999differential}, has received interest in recent years. In the game, the defenders are constrained to move along the perimeter and try to capture the intruders while the intruders aim to reach the perimeter without being captured by the defenders~\cite{shishika2020review}. A number of previous works have solved this game with engagements on a planar game space~\cite{shishika2018local, von2020guarding, chen2021optimal} to obtain optimal strategies for defenders and intruders. In the real world, the perimeters that defenders want to defend are not in 2D but in 3D. For instance, a perimeter of a building that defenders aim to protect can be enclosed by a hemisphere. Accordingly, the defender robots should be able to move in three-dimensional space. For example, aerial robots~\cite{lee2020experimental, nguyen2019mavnet, chen2020sloam, lee2016drone} have been well studied in various settings such as power plants~\cite{lee2020experimental}, penstocks~\cite{nguyen2019mavnet}, forests~\cite{chen2020sloam}, and disaster sites~\cite{lee2016drone}, and all these settings can be real-world use-cases for perimeter defense. For instance, an intruder attacks a military base in the forest and a defender aims to capture the intruder.

This work tackles the perimeter defense problem in a domain where multiple agents collaborate to accomplish a task. Multi-agent collaboration has been explored in many areas including rapid environmental mapping~\cite{thrun2000real}, search and rescue~\cite{baxter2007multi}, target tracking~\cite{zhou2018active, zhou2019sensor}, precision agriculture~\cite{kazmi2011adaptive}, and wireless networks~\cite{sharma2016uav}. Our approach employs a team of robots that work collectively towards a common goal of defending a perimeter, and we focus on developing decentralized strategies for the team of defenders for various reasons: (i) the teammates can be dynamically added or removed without disrupting explicit hierarchy; (ii) the centralized system may fail to cope with the high dimensionality of a team's joint state space; and (iii) the communication within the defender team at large scales is not guaranteed. For these reasons, we propose a framework where a team of defenders collaborates to defend the perimeter with decentralized strategies based on local perceptions. 


In this paper, we explore learning-based approaches to learn relevant policy by imitating expert algorithms such as the maximum matching~\cite{chen2014multiplayer}. Running the exhaustive search using the maximum matching algorithm is very expensive at large scales since this method is combinatorial in nature and assumes centralized information with full observability. We choose graph neural networks as the learning paradigm and demonstrate that the trained networks based on GNN can perform close to the expert policy. GNNs are the natural technique with the properties of decentralized communications that capture the neighboring interactions and transferability that allows for generalization to previously unseen scenarios~\cite{ruiz2021graph}. We demonstrate that our proposed GNN-based networks can generalize to large scales in solving the multi-robot perimeter defense. 

With this insight, we make the following primary contributions in this paper:
\begin{itemize} 
    \item \textbf{Framework for decentralized perimeter defense using graph neural networks.} We propose a novel framework that utilizes a graph-based representation of the perimeter defense game. To the best of our knowledge, we are the first to solve the decentralized hemisphere perimeter defense problem by learning decentralized strategies.
    
    \item \textbf{Robust perimeter defense performance with scalability.} We demonstrate that our methods perform close to an expert policy (i.e., maximum matching~\cite{chen2014multiplayer}) and are superior to other baseline algorithms. Our proposed networks are trained at a small scale and can generalize to large scales.
\end{itemize}

\section{Related Work}
\label{sec:related}

\textbf{Perimeter Defense:} In a perimeter defense game, defenders aim to capture intruders by moving along the perimeter while intruders try to reach the perimeter without being captured by defenders. We refer to~\cite{shishika2020review} for a detailed survey. Many previous works dealt with engagements on a planar game space~\cite{shishika2018local, von2020guarding, chen2021optimal, Macharet2020Adaptive}. For example, a cooperative multiplayer perimeter-defense game was solved on a planar game space in~\cite{shishika2018local}. In addition, guarding a circular target by patrolling its perimeter was considered in~\cite{von2020guarding}. Later, a formulation of the perimeter defense problem as an instance of the flow networks was proposed in~\cite{chen2021optimal}. Furthermore, an adaptive partitioning strategy based on intruder arrival estimation was proposed in~\cite{Macharet2020Adaptive}.  

High-dimensional extensions of the perimeter defense problem have been recently explored~\cite{lee2021guarding, yan2022matching, lee2020perimeter, lee2021defending}. Lee and Bakolas \cite{lee2021guarding} analyzed the two-player differential game of guarding a closed convex target set from an attacker in high-dimensional Euclidean spaces. Yan et al. \cite{yan2022matching} studied a 3D multiplayer reach-avoid game where multiple pursuers defend a goal region against multiple evaders. Lee et al. \cite{lee2020perimeter, lee2021defending} considered a game played between aerial defender and ground intruder.

All of the aforementioned works focus on solving centralized perimeter defense problems, which assume that players have global knowledge of other players' states. However, decentralized control becomes a necessity as we reach a large number of players. To remedy this problem, Velhal et al. \cite{velhal2022decentralized} formulated the perimeter defense game into a decentralized multi-robot Spatio-temporal multitask assignment problem on the perimeter of a convex shape. Paulos et al. \cite{paulos2019decentralization} proposed neural network architecture for training decentralized agent policies on the perimeter of a unit circle, where defenders have simple binary action spaces. Our work focuses on the high-dimensional perimeter, specialized to a hemisphere, with continuous action space. We solve multi-agent perimeter defense problems by learning decentralized strategies with graph neural networks.

\textbf{Graph Neural Networks:}
We leverage graph neural networks as the learning paradigm because of their desirable properties of decentralized architecture that captures the interactions between neighboring agents and transferability that allows for generalization to previously unseen cases~\cite{ruiz2021graph, Gama19-Architectures}. In addition, GNNs have shown great success in various multi-robot problems such as formation control~\cite{Tolstaya19-Flocking}, path planning~\cite{li2020message}, task allocation~\cite{wang2020learning}, and multi-target tracking~\cite{zhou2021graph}. Particularly, Tolstaya et al.~\cite{Tolstaya19-Flocking} utilized a GNN to learn a decentralized flocking behavior for a swarm of mobile robots by imitating a centralized flocking controller. Later, Li et al.~\cite{li2020message} implemented GNNs to find collision-free paths for multiple robots in obstacle-rich environments. They demonstrated their decentralized path planner achieves a near-expert performance with local observations and neighboring communication only, which can also generalize to larger networks of robots. The GNN-based approach was also employed to learn solutions to the combinatorial optimization problems in a multi-robot task scheduling scenario~\cite{wang2020learning} and multi-target tracking scenario~\cite{zhou2021graph}. 


\begin{figure}[!t]
\centering
\subfloat[]{\includegraphics[width=0.23\textwidth]{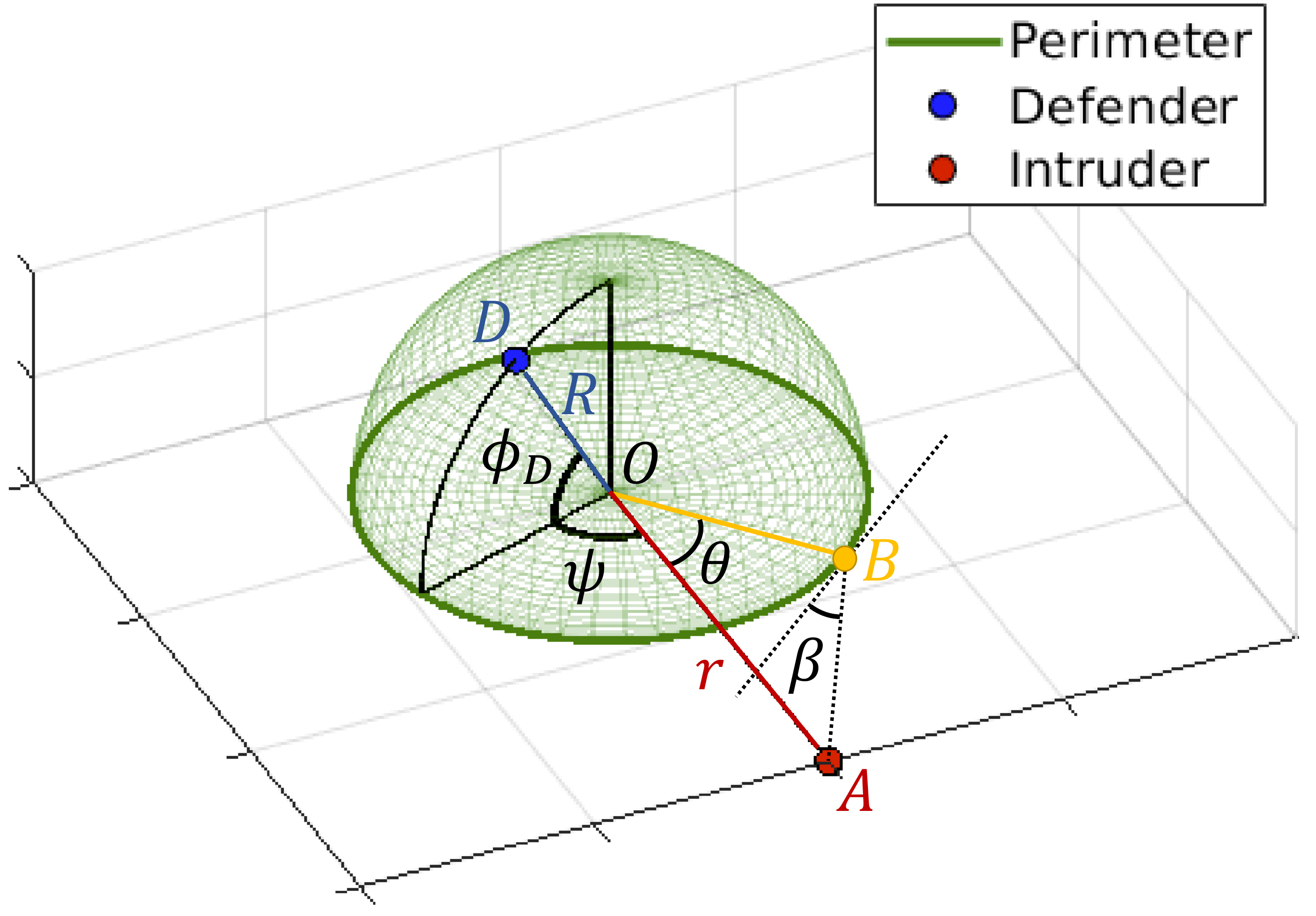}}
\hfil
\subfloat[]{\includegraphics[width=0.23\textwidth]{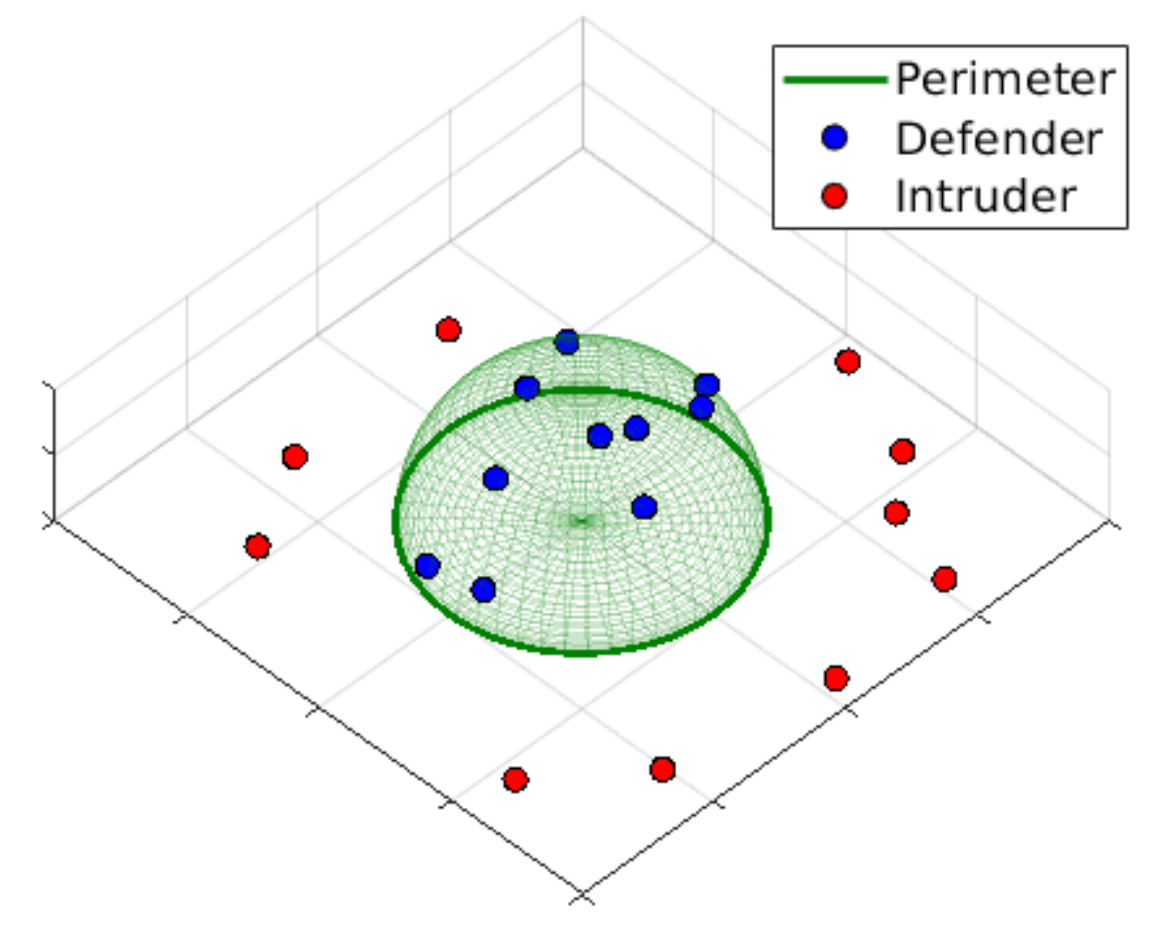}}
\caption{(a) Coordinates and relevant variables in the hemisphere defense game. (b) Instance of 10 vs. 10 perimeter defense.}
\label{fig:hemisphere}
\end{figure}

\begin{figure*}[b]
    \includegraphics[width=1\textwidth]{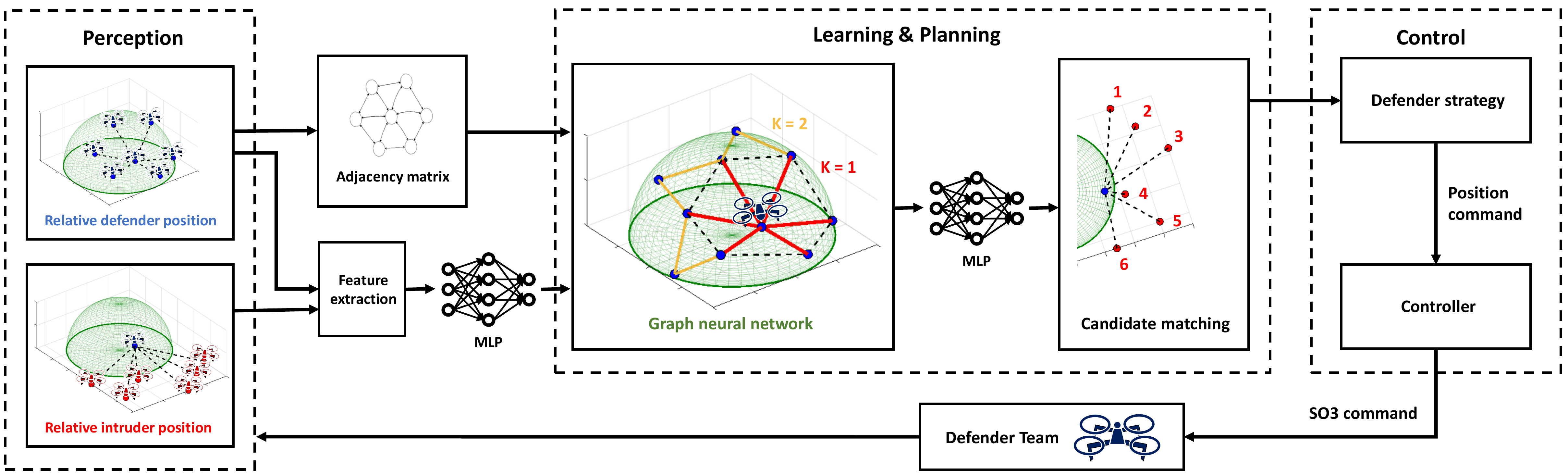}
    \caption{Overall framework. Perception module collects local information. Learning \& Planning module processes the collected information using GNN through $K$-hop neighboring communications. Control module computes the optimal strategies and executes the controller to close the loop.}
    \label{fig:framework}
\end{figure*}

\section{Problem Formulation}
\label{sec:problem}
We consider a hemispherical dome with radius of $R$ as perimeter (Fig.~\ref{fig:hemisphere}). The hemisphere constraint is for the defender to safely move around the perimeter (e.g. building). In this game, consider two sets of players: $\textbf{D}=\{D_i\}^N_{i=1}$ denoting $N$ defenders, and $\textbf{A}=\{A_j\}^N_{j=1}$ denoting $N$ intruders. A defender $D_i$ is constrained to move on the surface of the dome while an intruder $A_j$ is constrained to move on the ground plane. We will drop the indices $i$ and $j$ when they are irrelevant. An instance of 10 vs. 10 perimeter defense is shown on the right in Fig. \ref{fig:hemisphere}. The positions of the players in spherical coordinates are: $\zd=[\psi_D,\phi_D,R]$ and $\za=[\psi_A,0,r]$, where $\psi$ and $\phi$ are the azimuth and elevation angles, which gives the relative position as: $\mf z \triangleq [\psi,\phi,r]$, where $\psi\triangleq \psi_A-\psi_D$ and $\phi\triangleq \phi_D$. The positions of the players can also be described in Cartesian coordinates as: $\xd$ and $\xa$. All agents move at unit speed, defenders capture intruders by closing within a small distance $\epsilon$, and both defender and intruder are consumed during capture. An intruder wins if it reaches the perimeter (i.e., $r(t_f)=R$) at time $t_f$ without being captured by any defenders (i.e., $||\mf x_{A_i}(t) - \mf x_{D_j}(t)||>\epsilon, \forall D_j \in \mathbf{D}, \forall t < t_f$). A defender wins by capturing an intruder or preventing it from scoring indefinitely (i.e., $\phi(t)=\psi(t)=0$, $r(t)>R$). The main interest of this work is to maximize the number of captures by defenders, given a set of initial configurations.


To maximize the number of captures during $N$ vs. $N$ defense, we first recall the dynamics of a 1 vs. 1 perimeter defense game. Given $\zd$, $\za$, we call $\textit{breaching point}$ as a point on the perimeter at which the intruder tries to reach the target, as shown $B$ in Fig.~\ref{fig:hemisphere}(a). It is proved in~\cite{lee2020perimeter} that given the current positions of defender $\mf z_D$ and intruder $\mf z_A$, there exists a unique breaching point $B$ such that the optimal strategy for both defender and intruder is to move towards it, known as \textit{optimal breaching point}. We call the \textit{target time} as the time to reach $B$ and define $\tdd$ as the \textit{defender target time}, $\taa$ as the \textit{intruder target time}, and the following as \textit{payoff} function:

\bql
\pp = \tdd -\taa \label{eq:payoff}
\eql   

The defender reaches $B$ faster if $p<0$ and the intruder reaches $B$ faster if $p>0$. Thus, the defender aims to minimize $p$ while the intruder aims to maximize it. From this, we infer that maximizing the number of captures in $N$ vs. $N$ defense is the same as finding a matching between the defenders and intruders so that the number of the negative payoff of assigned pairs is maximized. In an optimal matching, the number of negative payoffs stays the same throughout the overall game since the optimality in each game of defender-intruder pairs is given as a \textit{Nash equilibrium} \cite{lee2020perimeter}.

The expert assignment policy is a \textit{maximum matching}~\cite{shishika2018local, chen2014multiplayer}. To execute this algorithm, we generate a bipartite graph with $\textbf{D}$ and $\textbf{A}$ as two sets of nodes (i.e., $\mc{V}=\{1,2,..,N\}$), and define the potential assignments between defenders and intruders as the edges. For each defender/node $D_i$ in $\textbf{D}$, we find all the intruders/nodes $A_j$ in $\textbf{A}$ that are sensible by the defender and compute the corresponding payoffs $p_{ij}$ for all the pairs. We say that $D_i$ is \textit{strongly assigned} to $A_j$ if $p_{ij}<0$. Using the edge set $\mathcal{E}$ given by maximum matching, we can maximize the number of strongly assigned pairs. For uniqueness, we choose a matching that minimizes the \textit{value of the game}, which is defined as
\begin{align}\label{eqn:V}
    V = \sum_{(D_i,A_j)\in \mathcal{E}^*} p_{ij},
\end{align}
where $\mathcal{E}^*$ is the subset of $\mathcal{E}$ with negative payoff (i.e. $\mathcal{E}^*= \{(D_i,A_j)\in\mathcal{E} \mid p_{ij}<0\}$). This unique assignment ensures that the number of captures is maximized at the earliest possible. However, running the exhaustive search using maximum matching algorithm can be very expensive as the team size increases. This method is combinatorial in nature and assumes centralized information with full observability. Instead, we aim to find decentralized strategies that uses local perceptions $\{\mc{Z}_i\}_{i\in\mc{V}}$ (see~\Cref{subsec:perception}). To this end, we formalize the main problem of this paper as follows.

\textit{Problem 1 (Decentralized Perimeter Defense with Graph Neural Networks):}
Design a GNN-based learning framework to learn a mapping $\mc{M}$ from the defenders' local perceptions $\{\mc{Z}_i\}_{i\in\mc{V}}$ and the communication graph $\mc{G}$ to the defenders' actions $\mc{U}$, i.e., $\mc{U} = \mc{M}(\{\mc{Z}_i\}_{i\in\mc{V}}, \mc{G})$, such that $\mc{U}$ is as close as possible to action set $\mc{U}^\texttt{g}$ selected by a centralized expert algorithm.  
\label{prob:learning}

We describe in detail our learning architecture for solving Problem 1 in the following section.

\section{Method}
\label{sec:method}

In this work, we learn decentralized strategies for perimeter defense with graph neural networks. Inference of our approach is in real-time, which is scalable to large scales. We use an expert assignment policy to train a team of defenders who share information through communication channels. In~\Cref{subsec:perception}, we introduce the perception module for processing the features that are input to GNN. Learning the decentralized algorithm through GNN and planning the candidate matching for the defenders are discussed in~\Cref{subsec:learnandplan}. The control of the defender team is explained in~\Cref{subsec:control}, and the training procedure is detailed in~\Cref{subsec:training}. The overall framework is shown in Fig.~\ref{fig:framework}. 

\subsection{Perception}\label{subsec:perception}
We consider $N$ aerial defenders that are tasked to perceive and intercept $N$ intruders on the ground. We assume that each defender $D_i$ is equipped with a sensor (e.g., a camera sensor) and faces outwards from the perimeter with a field of view \textit{FOV}. For each $i$, a defender observes the set of intruders $A_j$, and the relative positions in spherical coordinates between $D_i$ and $A_j$ are represented by $\mc{Z}_i^A = \{\mf z_{ij}^A\}_{j\in\mathit{N}_A^f}$ where $\mathit{N}_A^f$ is the number of intruder features. Each defender $D_i$ also communicates with nearby defenders $D_j$ within its communication range $r_c$. For each $i$, the relative positions between $D_i$ and $D_j$ are represented by $\mc{Z}_i^D = \{\mf z_{ij}^D\}_{j\in\mathit{N}_D^f}$ where $\mathit{N}_D^f$ is the number of defender features. Feature extraction is performed by concatenating the relative positions of observed intruders and communicated defenders, forming the local perceptions $\mc{Z}_i = \{\mc{Z}_i^A, \mc{Z}_i^D\}$. The extracted features are fed into a multi-layer perceptron (MLP) to generate the post-processed feature vector $\mf x_i$, which will be exchanged among neighbors through communications.

\subsection{Learning \& Planning}\label{subsec:learnandplan}
We implement graph neural networks with $K$-hop neighbors. All defenders communicate their perceived features with neighboring robots. The communication graph $\mc G$ is formed by connecting the nearby defenders within the communication range $r_c$, and the resulted adjacency matrix $\mf S$ is given to the graph neural networks. The output of the GNN, which represents the fused information from the $K$-hop communications, is then processed with another MLP to provide a candidate matching for each defender. The output from the multi-layer perceptron is an assignment likelihood $\mc L$, which presents the probabilities of $\mathit{N}_A^f$ intruder candidates' likelihood to be matched with the given defender. For instance, an expert assignment likelihood $L_i^g$ for $D_i$ in~\Cref{fig:framework} would be $[0,0,1,0,0,0]$ if the third intruder (i.e., $A_3$) is matched with $D_i$ by the expert policy (i.e., maximum matching). The planning module selects the intruder candidate $A_j$ so that the matching pair $(D_i, A_j)$ would resemble the expert policy with the highest probability. It is worth noting that our approach renders a decentralized assignment policy given that only neighboring information is exchanged.

\subsection{Control}\label{subsec:control}
This module handles all the matched pairs $(D_i, A_j)$ and computes the optimal breaching points for each of the one-on-one hemisphere perimeter defense games. The defender strategy module collectively outputs the position commands, which are towards the direction of the optimal breaching points. The SO(3) command~\cite{mellinger2011minimum} that consists of thrust and moment to control the robot at a low level is then passed to the defender team $\textbf{D}$ for control. The defenders move based on the commands to close the perception-action loop. Notably, the expert assignment likelihood $\mc L^g$ would result in the expert action set $\mc U^g$ (defined in~\Cref{prob:learning}).

\subsection{Training Procedure}\label{subsec:training}
To train our proposed networks, we use imitation learning to mimic an expert policy given by maximum matching, which provides the optimal assignment likelihood $\mc L^g$ given the defenders' local perceptions $\{\mc{Z}_i\}_{i\in\mc{V}}$ and the communication graph $\mc G$. The training set $\mc D$ is generated as a collection of these data: $\mc D = \{(\{\mc{Z}_i\}_{i\in\mc{V}}, \mc G, \mc L^g)\}$. We train the mapping $\mc M$ to minimize the cross-entropy loss between $\mc L^g$ and $\mc L$. We show that the trained $\mc M$ provides $\mc U$ that is close to $\mc U^g$. The number of learnable parameters in our networks is independent of the number of team sizes $N$. Therefore, we can train our networks on a small scale and generalize our model to large scales by learning decentralized strategies. 

Our model architecture consists of a 2-layer MLP with 16 and 8 hidden layers to generate the post-processed feature vector $\mf x_i$, a 2-layer GNN with 32 and 128 hidden layers to exchange the collected information from defenders, and a single-layer MLP to produce an assignment likelihood $\mc L$. The layers in MLP and GNN are followed by ReLU. We use the Adam optimizer with a momentum of 0.5. The learning rate is scheduled to decay from $5\times10^{-3}$ to $10^{-6}$ within 1500 epochs with batch size 64, using cosine annealing. We choose these hyperparameters for the best performance.


\section{Experiments}
\label{sec:experiments}
The experiments are conducted using a 12-core 3.50GHz i9-9920X CPU and an Nvidia GeForce RTX 2080 Ti GPU. We implement the proposed networks using PyTorch v1.10.1~\cite{paszke2019pytorch} accelerated with Cuda v10.2 APIs. 
The used parameters are summarized in~\Cref{tab:3}.

\begin{table}[t]
\centering
\caption{Parameter setup} 
\begin{tabular}{c | c c}
\hline
Parameter name & Symbol & Value\\
\hline
Capturing distance & $\epsilon$ & 0.02\\
Field of view & $FOV$ & $\pi$ \\
Number of intruder features & $\mathit{N}_A^f$ & 10\\
Number of defender features & $\mathit{N}_D^f$ & 3\\
Communication range & $r_c$ & 1\\
Default team size & $N_{def}$ & 10 \\
\hline
\end{tabular}
\label{tab:3}
\end{table}

\subsection{Datasets}\label{subsec:dataset}
We evaluate our decentralized networks using imitation learning where the expert assignment policy is the maximum matching. The perimeter is a hemisphere with a radius $R$, which is defined by $R = \sqrt{N/N_{def}}$ where $N$ is team size and $N_{def}$ is a default team size. Since running the maximum matching is very expensive at large scales (e.g. $N>10$), we set the default team size $N_{def}=10$. In this way, $R$ represents the scale of the game; for instance when $N=40$, $R$ becomes 2, which indicates that the scale of the problem's setting is doubled compared to the setting when $R=1$. Given the team size $N=10$, our experimental arena has a dimension of $10\times 10\times 1$ m. We randomly sample 10 million examples of defender's local perception $\mc Z_i$ and find corresponding $\mc G$ and $\mc L^g$ to prepare the dataset, which is divided into a training set (60\%), a validation set (20\%), and a testing set (20\%).

\subsection{Metrics}\label{subsec:metrics}
We are mainly interested in the percentage of intruders caught (i.e., number of captures/total number of intruders). At small scales (e.g. $N\leq 10$), an expert policy (i.e., maximum matching) can be run and a direct comparison between the expert policy and our policy is available. At large scales (e.g. $N>10$), maximum matching is too expensive to run. Thus we compare our algorithm with other baseline approaches: \textit{greedy}, \textit{random}, and \textit{mlp}, which will be explained in~\Cref{subsec:compared}. To observe the scalability between small and large scales, we run five different algorithms for each scale: \textit{expert}, \textit{gnn}, \textit{greedy}, \textit{random}, and \textit{mlp}. In all cases, we compute the \textit{absolute accuracy}, which is defined by the number of captures divided by the team size, to verify if our network can generalize to any team size. Furthermore, we also calculate the \textit{comparative accuracy}, defined as the ratio of the number of captures by \textit{gnn} to the number of captures by another algorithm, to observe comparative results.

\subsection{Compared Algorithms}\label{subsec:compared}

\paragraph{Greedy} The greedy algorithm can be run in polynomial time and thus becomes a good candidate algorithm to be compared with our approach using GNN. For a fair comparison, we run a decentralized greedy algorithm based on local perception $\mc Z_i$ given $D_i$. We enable $K$-hop neighboring communications so that the sensible region of a defender is expanded as if the networking channels of GNN are active. The defender $D_i$ computes the payoff $p_{ij}$ based on any sensible intruder $A_j$ and greedily chooses an assignment that minimizes the payoff $p_{ij}$. 

\paragraph{Random} The random algorithm is similar to the greedy algorithm in that the $K$-hop neighboring communications are enabled for the expanded perception. Among sensible intruders, a defender $D_i$ randomly picks an intruder to determine the assignment.

\paragraph{MLP} We only train the current MLP of our proposed framework in isolation by excluding the GNN module. By comparing our GNN framework to this algorithm, we may observe if the GNN gives any improvement.

\subsection{Results}\label{subsec:results}
\begin{figure}[b]
    \centering   
    \subfloat[6 vs. 6]{\includegraphics[width=0.23\textwidth]{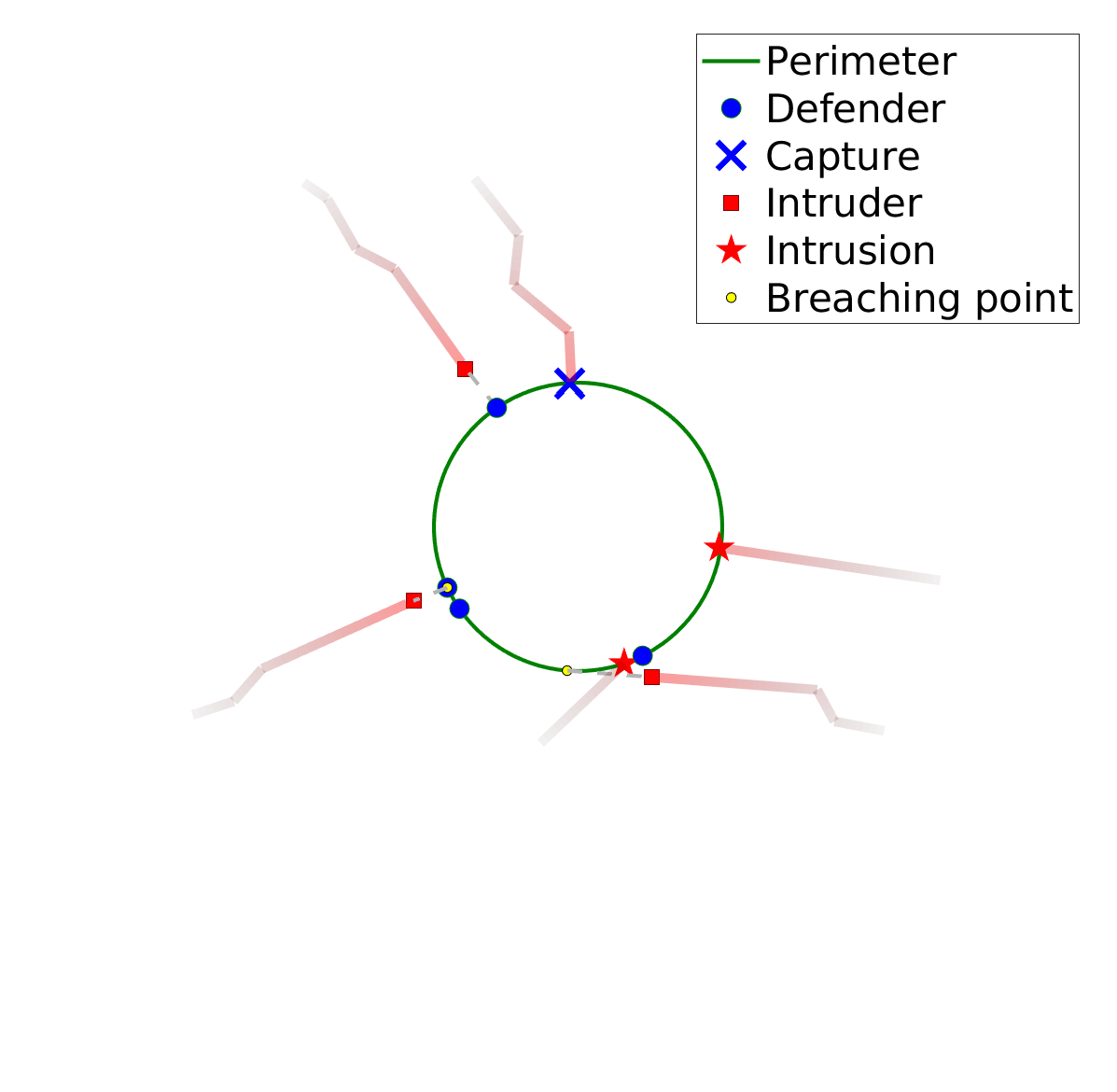}}
    \hfill
    \subfloat[100 vs. 100]{\includegraphics[width=0.23\textwidth]{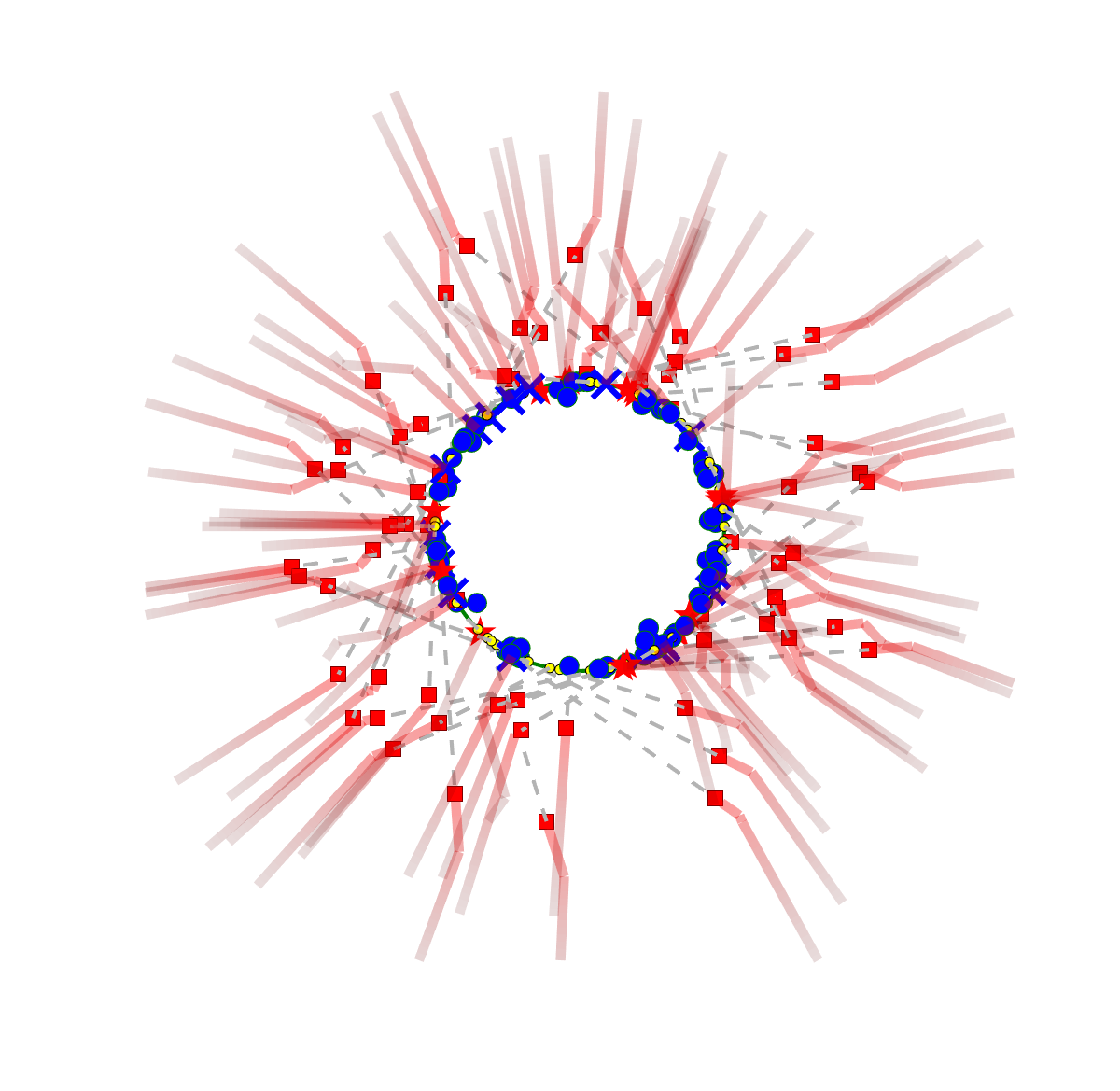}}
    \caption{Snapshots of simulated perimeter defense in top view using the proposed method \textit{gnn}. (a) 6 vs. 6 game. (b) 100 vs. 100 game.}
    \label{fig:game}
\end{figure}

We run the perimeter defense game in diverse scenarios with different team sizes and initial configurations to evaluate the performance of the learned networks. In large, we conduct the experiments at small ($N\leq 10$) and large ($N>10$) scales. The snapshots of the simulated perimeter defense game in top view with our proposed networks for different team sizes are shown in Fig.~\ref{fig:game}. The perimeter, defender state, intruder state, and breaching point are marked in green, blue, red, and yellow, respectively. We can observe that intruders try to reach the perimeter. Given the defender-intruder matches, the intruders execute their respective optimal strategies to move towards the optimal breaching points. If an intruder successfully reaches it without being captured by any defender, the intruder is consumed and leaves a marker called ``Intrusion". If an intruder fails and is intercepted by a defender, both agents are consumed and leave a marker called ``Capture". The points on the perimeter aimed by intruders are marked as ``Breaching point". In all runs, the game ends at \textit{terminal time} $T_f$ when all the intruders are consumed. See the supplemental video for more results.

\begin{figure}[t!]
    \centering  
    \subfloat[\textit{gnn}]{\includegraphics[width=0.2\textwidth]{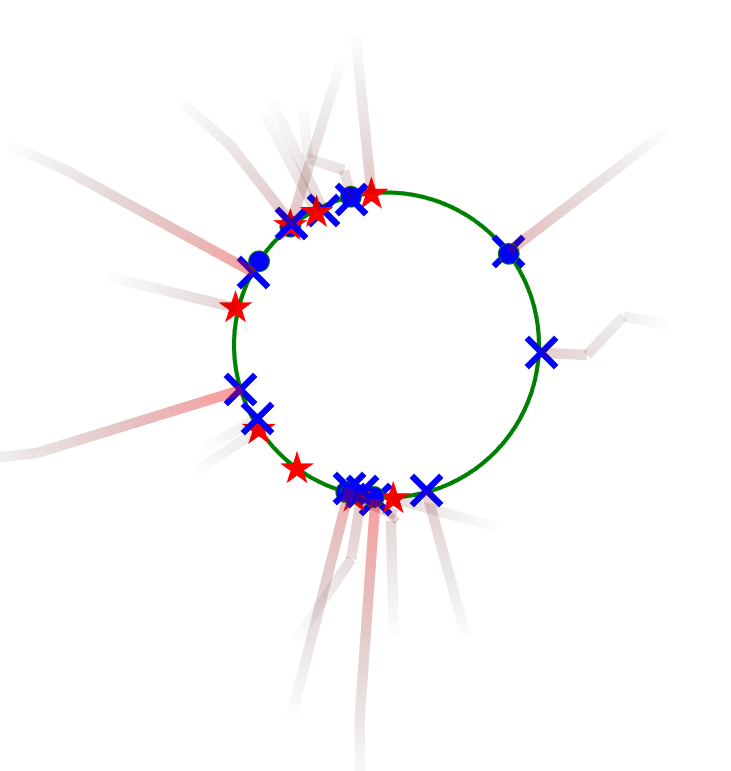}}
    \subfloat[\textit{greedy}]{\includegraphics[width=0.2\textwidth]{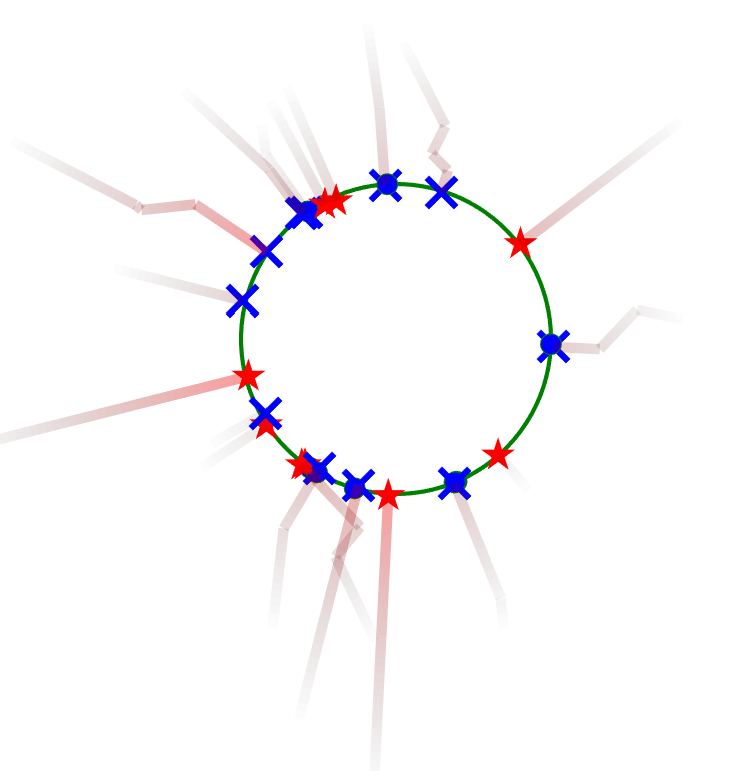}}
    \hfill
    \subfloat[\textit{random}]{\includegraphics[width=0.2\textwidth]{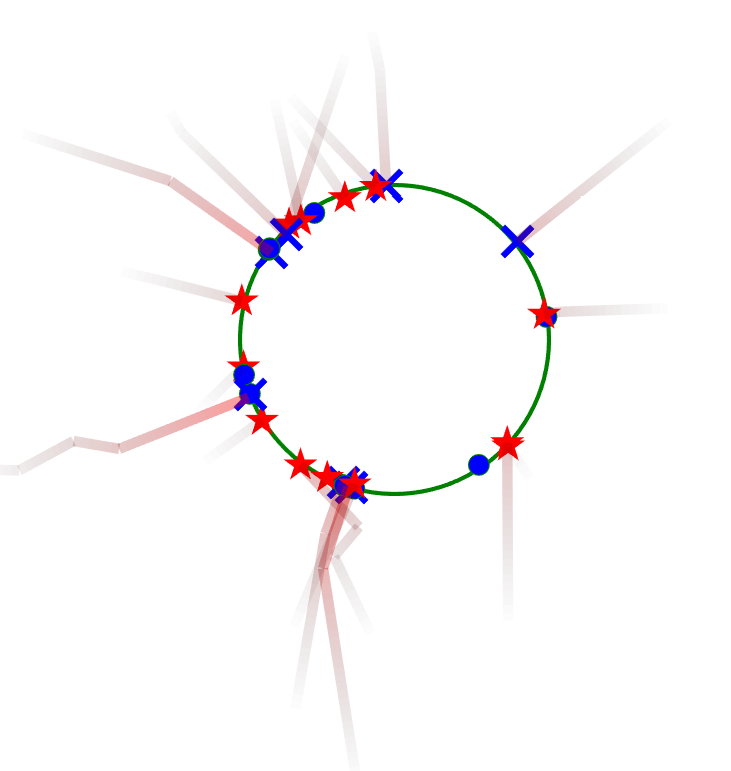}}
    \subfloat[\textit{mlp}]{\includegraphics[width=0.2\textwidth]{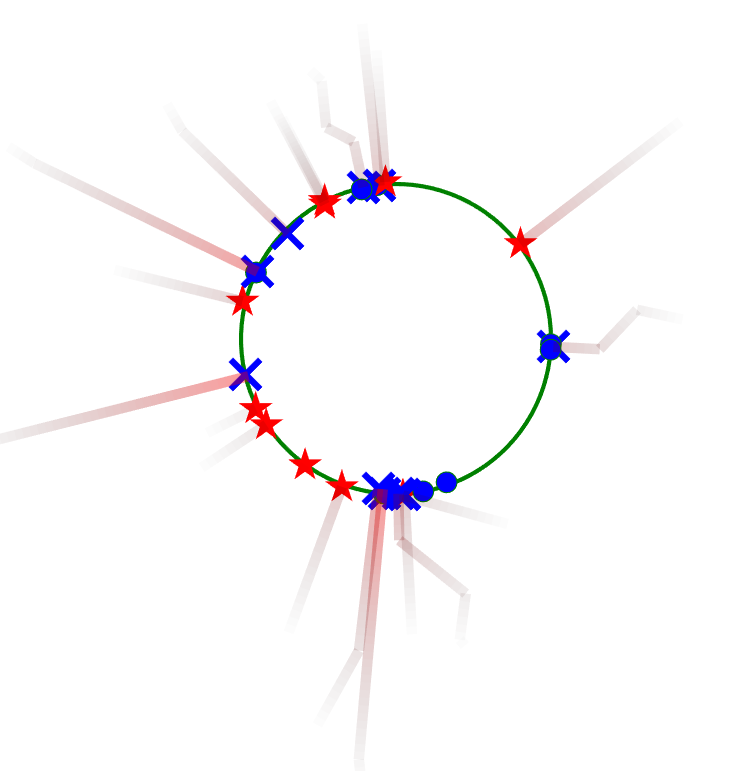}}
    \caption{Snapshots of simulated 20 vs. 20 perimeter defense in top view at terminal time $T_f$ using the four algorithms. The number of captures using these algorithms are 12, 11, 10, and 7, respectively. (a) \textit{gnn} algorithm. (b) \textit{greedy} algorithm. (c) \textit{random} algorithm. (d) \textit{mlp} algorithm.}
    \label{fig:endgame}
\end{figure}

As mentioned in~\Cref{subsec:dataset}, we run the five algorithms \textit{expert}, \textit{gnn}, \textit{greedy}, \textit{random}, and \textit{mlp} at small scales, and run \textit{gnn}, \textit{greedy}, \textit{random}, and \textit{mlp} in large scales. As an instance, the snapshots of simulated 20 vs. 20 perimeter defense game in top view at terminal time $T_f$ using the four algorithms are displayed in Fig.~\ref{fig:endgame}. The four subfigures (a)-(d) show that these algorithms exhibit different performance although the game begins with the same initial configuration in all cases. The number of captures by these algorithms \textit{gnn}, \textit{greedy}, \textit{random}, and \textit{mlp} are 12, 11, 10, 7, respectively.

\begin{figure}[t!]
    \centering
    \includegraphics[width=0.3899\textwidth]{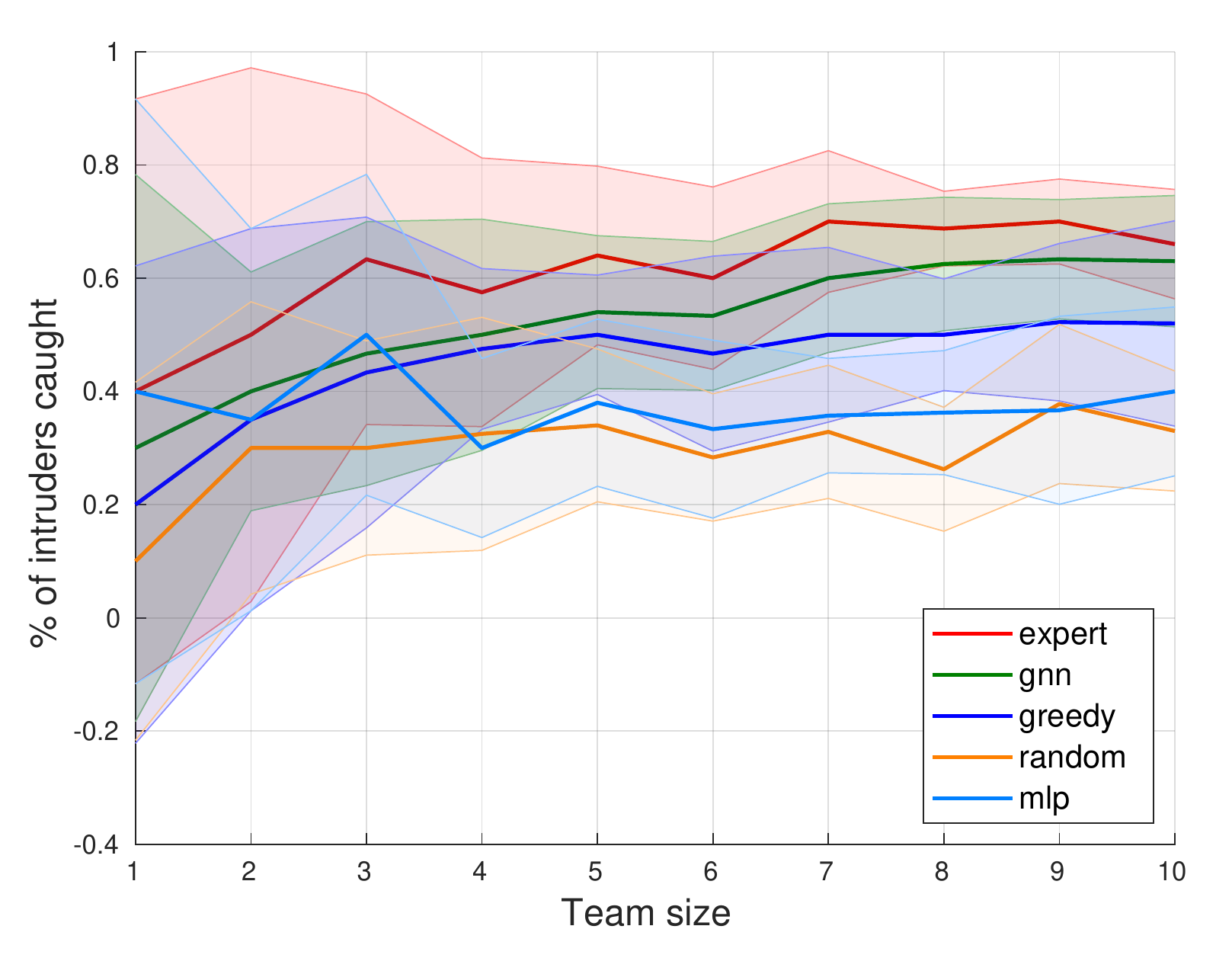}
    \includegraphics[width=0.3899\textwidth]{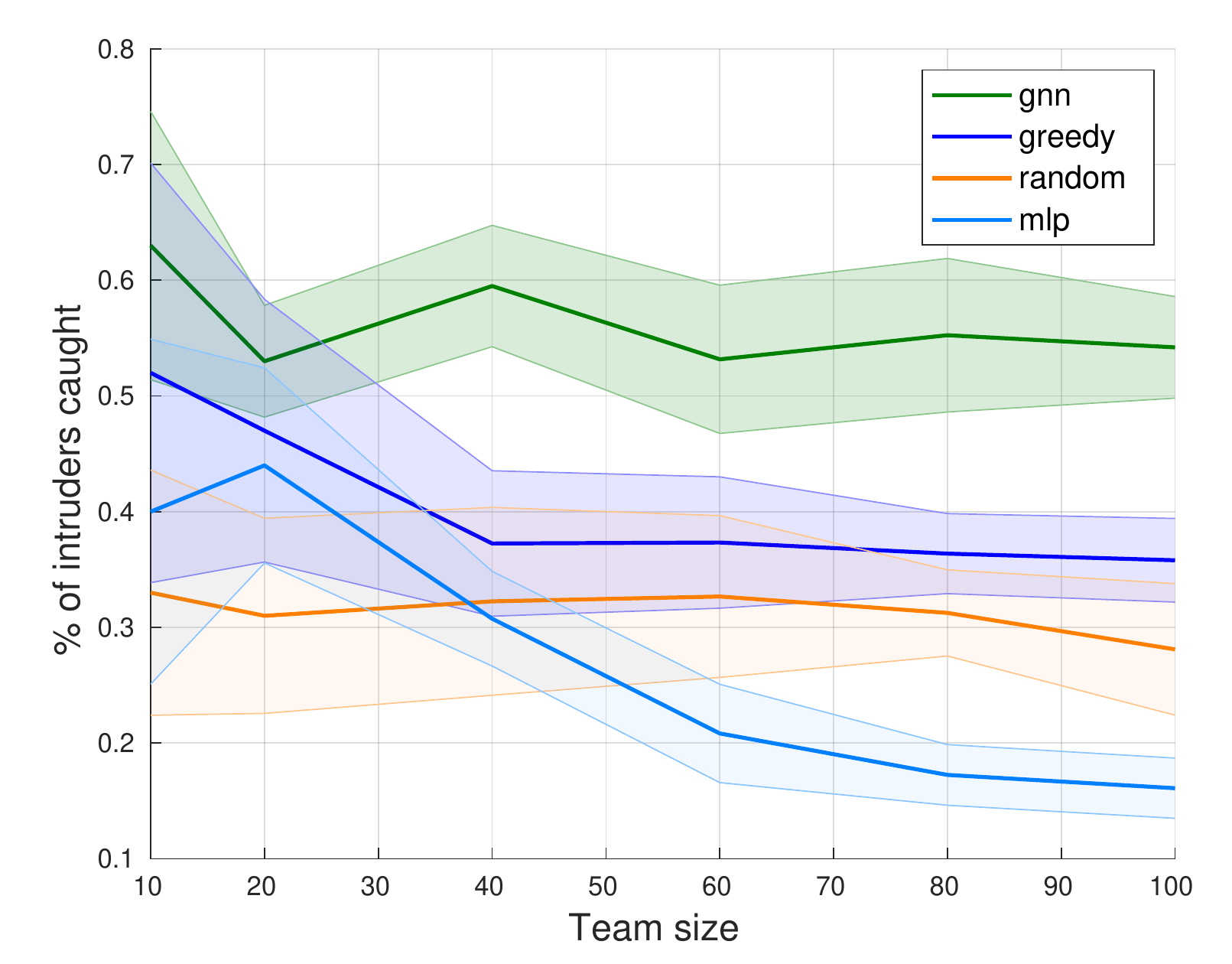}
    \caption{Percentage of intruders caught (average and standard deviation over 10 trials) by algorithms on small ($N\leq10$) and large ($N>10$) scales.}
    \label{fig:scale}
\end{figure}

The overall results of the percentage of intruders caught by each of these methods are depicted in Fig.~\ref{fig:scale}. It is observed that \textit{gnn} outperforms other baselines in all cases, and performs close to \textit{expert} at the small scales. In particular, given that our default team size $N_{def}$ is 10, the performance of our proposed algorithm stays competitive with that of the expert policy near $N=10$.

At large scales, the percentage of captures by \textit{gnn} stays constant, which indicates that the trained network generalizes to the large scales even if the training has been performed at the small scale. The percentage of captures by \textit{greedy} also seems constant but performs much worse than \textit{gnn} as the team size gets large. At small scales, only a few combinations are available in matching defender-intruder pairs and thus the \textit{greedy} algorithm would perform similarly to the expert algorithm. As the number of agents increases, the number of possible matching increases exponentially so the \textit{greedy} algorithm performs worse since the problem complexity gets much higher. The \textit{random} approach performs worse than all other algorithms at small scales, but the \textit{mlp} begins to perform worse than the \textit{random} when the team size increases over 40. This tendency tells that the policy trained only with MLP cannot be scalable at large scales. Since the training is done with 10 agents, it is optimal near $N=10$, but the \textit{mlp} cannot work at larger scales and even performs worse than the \textit{random} algorithm. It is confirmed that the GNN added to the MLP significantly improves the performance.

\begin{table}[b]
\centering
\caption{Accuracy for small scales} 
\begin{tabular}{c | c c c c c}
\hline
Team Size & 2 & 4 & 6 & 8 & 10\\
\hline
Absolute accuracy & 0.40 & 0.50 & 0.53 & 0.63 & 0.63\\
gnn vs. expert & 0.80 & 0.87 & 0.89 & 0.91 & 0.95\\
gnn vs. greedy & 1.14 & 1.05 & 1.14 & 1.25 & 1.21\\
gnn vs. random & 1.33 & 1.54 & 1.88 & 2.38 & 1.91\\
gnn vs. mlp & 1.14 & 1.67 & 1.60 & 1.72 & 1.58\\
\hline
\end{tabular}
\label{tab:1}
\end{table}

\begin{table}[b]
\centering
\caption{Accuracy for large scales}
\begin{tabular}{c | c c c c c}
\hline
Team Size & 20 & 40 & 60 & 80 & 100\\
\hline
Absolute accuracy & 0.53 & 0.59 & 0.53 & 0.55 & 0.54\\
gnn vs. greedy & 1.13 & 1.59 & 1.42 & 1.52 & 1.51\\
gnn vs. random & 1.71 & 1.85 & 1.63 & 1.77 & 1.93\\
gnn vs. mlp & 1.20 & 1.94 & 2.55 & 3.20 & 3.37\\
\hline
\end{tabular}
\label{tab:2}
\end{table}

To quantitatively evaluate the proposed method, we report the \textit{absolute accuracy} and \textit{comparative accuracy} in~\Cref{tab:1} and~\Cref{tab:2}. As expected, the absolute accuracy reaches the maximum when team size approaches $N=10$. The overall values of the absolute accuracy are fairly consistent except for when $N=2$. We conjecture that there may not be much information shared by the two defenders and there could be no sensible intruders at all based on initial configurations.

The comparative accuracy between \textit{gnn} and \textit{expert} shows that our trained policy gets much closer to the expert policy as $N$ approaches 10, and we expect the performance of \textit{gnn} to be close to that of \textit{expert} even at the large scales. The comparative accuracy between \textit{gnn} and other baselines shows that our trained networks perform much better than baseline algorithms at the large scales ($N\geq 40$) with an average of 1.5 times more captures. The comparative accuracy between \textit{gnn} and \textit{random} is somewhat noisy throughout the team size due to the nature of randomness, but we observe that our policy can outperform random policy with an average of 1.8 times more captures at small and large scales. We observe that \textit{mlp} performs worse than other algorithms at large scales.

Based on the comparisons, we demonstrate that our proposed networks, which are trained at a small scale, can generalize to large scales. Intuitively, one may think that \textit{greedy} would perform the best in a decentralized setting since each defender does its best to minimize the \textit{value of the game} (defined in~\Cref{eqn:V}). However, we can infer that  \textit{greedy}  does not know the intention of nearby defenders (e.g. which intruders to capture) so it cannot achieve the performance close to the centralized expert algorithm. Our method implements graph neural networks to exchange the information of nearby defenders, which perceive their local features, to plan the final actions of the defender team; therefore, implicit information of where the nearby defenders are likely to move is transmitted to each neighboring defender. Since the centralized expert policy knows all the intentions of defenders, our GNN-based policy learns the intention through communication channels. The collaboration among the defender team is the key for our \textit{gnn} to outperform \textit{greedy} approach. These results validate that the GNNs are ideal for our problem with the decentralized communication that captures the neighboring interactions and transferability that allows for generalization to unseen scenarios.


\section{Conclusion}
\label{sec:conclusion}

This paper proposes a novel framework that employs graph neural networks to solve the decentralized multi-agent perimeter defense problem. Our learning framework takes the defenders' local perceptions and the communication graph as inputs and returns actions to maximize the number of captures for the defender team. We train deep networks supervised by an expert policy based on maximum matching. To validate the proposed method, we run the perimeter defense game in different team sizes using five different algorithms: \textit{expert}, \textit{gnn}, \textit{greedy}, \textit{random}, and \textit{mlp}. Based on the comparison of the algorithms, we demonstrate that our GNN-based policy stays closer to the expert policy at small scales and the trained networks can generalize to large scales. One future work is to implement vision-based local sensing for the perception module, which would relax the assumptions of perfect state estimation. Realizing multi-agent perimeter defense with vision-based perception and communication within the defenders will be an end goal. Another future research direction is to leverage GNNs to learn resilient perimeter defense strategies against adversarial attacks that can compromise defenders' perceptions and communications~\cite{zhou2018resilient,zhou2021robust}.

\bibliography{reference}
\bibliographystyle{IEEEtran}

\end{document}